\documentclass[preprint,aps,prd,superscriptaddress,nofootinbib]{revtex4}%

\usepackage{amsmath}
\usepackage{amssymb}
\usepackage{graphicx}
\usepackage{dcolumn}
\usepackage{bm}
\usepackage{hyperref}
\usepackage{epsfig}
\usepackage{mathrsfs}

\begin{document}
\preprint{TUM-EFT 134/19}

\title{New method for fitting coefficients in standard model effective 
theory}
\author{Geoffrey~T.~Bodwin}
\email[]{gtb@anl.gov}
\affiliation{High Energy Physics Division, Argonne National Laboratory,
Argonne, Illinois 60439, USA}
\author{Hee~Sok~Chung}
\email[]{heesok.chung@tum.de}
\affiliation{Physik-Department, Technische Universit\"at M\"unchen,
James-Franck-Str.\ 1, 85748 Garching, Germany}
\affiliation{Excellence Cluster ORIGINS,
Boltzmannstrasse 2, D-85748 Garching, Germany}

\date{\today}
\begin{abstract}
We present an alternative method for carrying out a principal-component
analysis of Wilson coefficients in standard model effective field theory
(SMEFT).  The method is based on singular-value decomposition
(SVD). The SVD method provides information about the sensitivity of
experimental observables to physics beyond the standard model that
is not accessible in the Fisher-information method. In principle,
the SVD method can also have computational advantages over
diagonalization of the Fisher information matrix.  We demonstrate the
SVD method by applying it to the dimension-6 coefficients for the
process of top-quark decay to a $b$ quark and a $W$ boson and use this
example to illustrate some pitfalls in widely used fitting procedures.
We also outline an iterative procedure for applying the SVD method to
dimension-8 SMEFT coefficients.
\end{abstract}
\maketitle

\section{Introduction}

In recent years, standard model effective theory (SMEFT)
\cite{Buchmuller:1985jz,Hagiwara:1993ck,Giudice:2007fh,Grzadkowski:2010es}
has been a focus of activity in both the
theoretical and experimental particle-physics communities.  SMEFT has
been advocated as a means to quantify systematically deviations of the
global set of experimental measurements from the predictions of the
standard model. To this end, a number of efforts have been undertaken to
perform global fits of the Wilson coefficients of SMEFT operators. 
Some examples of global fits are contained in Refs.~\cite{Han:2004az,
Pomarol:2013zra,Chen:2013kfa,Ellis:2014dva,Wells:2014pga,Falkowski:2014tna,
deBlas:2017wmn,deBlas:2018tjm,Ellis:2018gqa,Aebischer:2018iyb,
Almeida:2018cld,Biekotter:2018rhp,Hartland:2019bjb,Cepeda:2019klc,
Brivio:2019ius}.

Several difficulties can arise in using data to constrain coefficients
in SMEFT. First, there are many operators in SMEFT (59 in
dimension~6) and potentially many data points that could be used in
fitting the coefficients of these operators. Second, theoretical
expressions for the SMEFT contributions to a given set of experimental
observables can contain ``flat directions'' (or nearly flat
directions) in the space of SMEFT coefficients, that is, directions
for which the observables are insensitive to the values of SMEFT
coefficients. Third, for analyses involving a limited sector of
observables, there may be fewer observables than SMEFT coefficients.
This situation will necessarily result in the existence of exactly flat
directions.

All of these difficulties can pose computational problems in
fitting SMEFT coefficients to data. A global fit may be
computationally challenging because standard methods for carrying out
fits may bog down when the number of observables and coefficients is
large.  When the number of coefficients is greater than the number of
observables and/or there are flat directions, methods of fitting that
minimize $\chi^2$ numerically may not converge reliably. In addition to
these technical issues, there is also an issue of principle: When there
are more coefficients than observables and/or flat directions, the
uncertainties in the coefficients can be highly correlated, and
bounds on values of individual SMEFT
coefficients may be very misleading.

This last issue of principle can be addressed by carrying out a
principal-component analysis (PCA) of the SMEFT coefficients. One way to
do this is by finding the eigenvalues and eigenvectors of the Fisher
information matrix, which, in the case of Gaussian statistics, is the
inverse of the correlation matrix~\cite{Berthier:2015gja,Berthier:2016tkq,Brivio:2017bnu,Brehmer:2017lrt,Brehmer:2019xox,Aoude:2020dwv}.
Since the Fisher matrix is nonsingular, diagonalization of the Fisher
matrix evades the computational problems that arise, when there
are flat directions, in minimizing $\chi^2$ numerically.

Another approach is to regularize $\chi^2$, so that it is nonsingular
and can be minimized by numerical methods~\cite{Murphy:2017omb}. In this
approach, the regulator could potentially introduce biases into the fit.

In this paper, we present an alternative method for carrying out the PCA
of the SMEFT coefficients that is based on singular-value decomposition
(SVD). The SVD method has the same advantages as the diagonalization of
the Fisher matrix in evading the computational problems that can arise
in numerical minimization of $\chi^2$. In addition, the SVD method
provides information about the sensitivity of experimental
observables to physics beyond the standard model that is not
accessible in the Fisher-information method.

There may also be algorithmic advantages in using SVD.  Many
well developed numerical methods exist both for diagonalization and for
SVD.  However, because the Fisher matrix is quadratic in the matrix to
which SVD is applied, its condition number is the square of the
condition number for the SVD matrix, and so, the precision may be worse
for diagonalization of the Fisher matrix than for
SVD~\cite{kalman-1996}.  Furthermore, if the number of SMEFT
coefficients is much larger than the number of observables, then the
matrix that is analyzed in the SVD approach is much smaller than the
Fisher matrix, and so the computation time may be smaller in the SVD
approach.  It remains to be seen whether these advantages will be
significant in extensive and/or iterated global fits of SVD
coefficients. 

We demonstrate the SVD method by applying it to a restricted class of
observables that appear in decay of the top quark to a $b$ quark and a
$W$ boson.\footnote{We note that global fits of SMEFT coefficients in
  the top-quark sector have been carried out in
  Refs.~\cite{Buckley:2015lku,Cirigliano:2016nyn,AguilarSaavedra:2018nen,Hartland:2019bjb}.}
This example allows us to show how the SVD method can be used to deal
with correlated theoretical and experimental errors and with the
difficulties of flat directions.

In fits of SMEFT coefficients in the literature, two methods that are
often employed are (1) fits in which all of the SMEFT coefficients but
one are set to zero and (2) fits in which all of the SMEFT coefficients
but one are marginalized (profiled). It has been emphasized in
Ref.~\cite{Berthier:2015gja} that both of these approaches are
misleading and are obviated by PCA. In this paper, we demonstrate in
explicit examples involving the SMEFT coefficients in the top-quark
sector, that approach (1) leads to overly optimistic constraints on the
SMEFT coefficients, while approach (2) leads to overly pessimistic
constraints on the SMEFT coefficients. Our ten-coefficient 
PCA results for the top-quark-decay SMEFT coefficients should be considered
to supersede the one-coefficient fit results in 
Ref.~\cite{Boughezal:2019xpp}, in which
approach (1) was used,
and the two-coefficient fits in Ref.~\cite{Boughezal:2019xpp}, which do not account completely for the high degree of
correlation between the uncertainties in the coefficients.

While we have not specifically demonstrated the utility of the SVD
method for a situation in which there is a large number of observables
and a large number of SMEFT coefficients, we are confident that it would
work reliably and efficiently in such a situation because of experience
with an application of SVD, in a different context, that involved the
fitting of thousands of data points with hundreds of coefficients
\cite{Bodwin:2018cwb}.

The SVD method that we present is based on a Gaussian uncertainty analysis.
While this, of course, is not completely general, it should prove to be
adequate at least for initial exploratory studies of the bounds on SMEFT
coefficients. The SVD method also requires that the observables depend
linearly on the SMEFT coefficients. This is the case in a computation
at leading order in the effective-field-theory expansion. As we will
describe later, the method can also be used iteratively to carry out a
PCA of the SMEFT coefficients in the case in
which higher-order contributions in the effective-field-theory expansion
are considered---provided that the expansion itself converges.

The remainder of this paper is organized as follows. In
Sec.~\ref{sec:SVD-method}, we outline the basics of SVD and present a
method for using SVD to carry out analyses of SMEFT
coefficients. Section~\ref{sec:top-decay} contains an illustration of
the use of SVD analysis in top-quark decay to a $b$ quark and a
$W$ boson. Here, we present examples of fits involving flat directions
and various numbers of SMEFT coefficients, and we contrast the results
from the PCA with those from the traditional fitting approaches (1) and
(2) that are mentioned above. In Sec.~\ref{sec:extension}, we discuss
the extension of the SVD approach to situations in which the
theoretical expressions for the observables depend nonlinearly on the
SMEFT coefficients.  Finally, in Sec.~\ref{sec:summary}, we summarize
our results.

\section{Singular-value decomposition method\label{sec:SVD-method}}
\subsection{Singular-value decomposition\label{sec:SVD}}

The singular-value-decomposition theorem states that an $m\times n$ 
matrix $M$ that contains either real or complex entries can always be 
decomposed as \cite{lawson-hanson-1995}
\begin{equation}
M=UWV^\dagger,
\label{eq:SVD}
\end{equation}
where $U$ is an $m\times m$ unitary matrix, $V$ is an $n\times n$
unitary matrix, $W$ is an $m\times n$ diagonal matrix with nonnegative
real numbers on the diagonals, and $\dagger$ denotes the Hermitian
conjugate (complex conjugate transpose). The diagonal entries of $W$ are
called the singular values. The matrix $W$ is unique, but the matrices
$U$ and $V$ are not. If $M$ is a square matrix, then $U$ and $V$ are
unique, up to phases that multiply each row of $V^\dagger$ and
corresponding inverse phases that multiply each column of $U$.  If $M$
is not a square matrix, additional ambiguities in $U$ and $V$ can
arise. Efficient computer codes exist for carrying out the SVD
decomposition of large matrices numerically. See, for example,          
Refs.~\cite{mathematica-11,numerical-recipes,python,gsl}.

SVD has the important property that it can be used to solve the linear 
least-squares problem, as we will now explain. Suppose that 
$M$ is an $m\times n$ matrix, $C$ is an $n$-dimensional column 
vector, and ${O}$ is an $m$-dimensional column   
vector. Further suppose that we wish to minimize 
\begin{equation}
(MC-{O})^2
\end{equation}
with respect to the elements of $C$ (coefficients). Here, the square 
denotes the inner product of $MC-{O}$ with itself. The solution of 
this problem is given by \cite{bjerhammar-1951}
\begin{equation}
\bar{C}=VW^{-1}U^\dagger{O},
\label{eq:SVD-solution}
\end{equation}
where $U$, $V$, and $W$ are the matrices that appear in the SVD
decomposition of $M$ [Eq.~(\ref{eq:SVD})] and $W^{-1}$ is the
Moore-Penrose pseudoinverse of $W$
\cite{moore-1920,bjerhammar-1951,penrose-1955,penrose-1956}, which is
obtained by taking the transpose of $W$ and replacing the nonzero
elements of $W^T$ with their inverses, while leaving the zero
elements unchanged.\footnote{Throughout this paper, when we refer to the
  pseudoinverse of a matrix, we mean the Moore-Penrose pseudoinverse,
  and we use the $-1$ power of a noninvertible matrix to denote its
  Moore-Penrose pseudoinverse.} Note that the solution in
Eq.~(\ref{eq:SVD-solution}) exists even if the matrix $M$ is
noninvertible. The solution exists, for example, if $M$ is not a square
matrix or if $M$ has a vanishing determinant.

The matrix $V^\dagger$ takes the elements of $C$ from the original basis
of coefficients to the principal-component basis. Each element of
$P^C=V^\dagger C$ is one of the principal components. Similarly, the
    matrix $U^\dagger$ yields the principal components of ${O}$:
$P^{O}=U^\dagger O$.
Owing to the phase
ambiguities in the SVD in each row in $V^\dagger$ and each row of
$U^\dagger$, each principal
component is defined only up to an overall phase. By virtue of the
unitarity of $V$ and of $U$, the principal components $P^C$ are
orthogonal to each other, and the principal components $P^{O}$ are
orthogonal to each other. Owing to the fact the $W$ is diagonal, each
principal component in $P^C$ is coupled to only a single principal
component in $P^{O}$, and {\it vice versa}.

The best-fit values of the principal components $P^C$ are given by the
elements of
\begin{equation}
\bar{P}^C=V^\dagger \bar{C}=W^{-1}U^\dagger{O},
\label{eq:best-fit-values}
\end{equation}
where we have used Eq.~(\ref{eq:SVD-solution}).  The fluctuation in $MC$
that is produced by a unit fluctuation in a principal component in
$P^C$ is given by the corresponding singular value in $W$.  Hence,
the uncertainties $\Delta P^C$ in the best-fit values
$\bar{P}^C$ are given by the inverses of the diagonal values
of $W$:
\begin{equation}
\Delta P^C_i=W_{ii}^{-1}.
\label{eq:uncertainty-formula}
\end{equation}
Because $W$ is diagonal, these uncertainties are uncorrelated.
Furthermore, they depend on the uncertainty of only a single
principal component in $P^{O}$.  The principal components that are
ill constrained because of the existence of flat directions correspond
to near-vanishing diagonal values of $W$.  That is, the SVD sequesters
linear combinations of coefficients that are ill constrained because
of the presence of flat directions in the coefficient space.

\subsection{Application of singular-value decomposition 
to the fitting of SMEFT coefficients}

 The fitting of the SMEFT coefficients is carried out by minimizing the
$\chi^2$, which is defined by 
\begin{equation}
\label{eq:chisq}%
\chi^2 = (O^{\rm SMEFT}-O^{\rm exp})^T (\sigma^2)^{-1}
(O^{\rm SMEFT}-O^{\rm exp}),
\end{equation}
where $O^{\rm exp}$ is the $N_{\rm obs}$-dimensional column vector of
experimental observables, $O^{\rm SMEFT}$ is the $N_{\rm
obs}$-dimensional column vector of theoretical predictions for the
observables in the SMEFT, $\sigma^2$ is the $N_{\rm obs}\times
N_{\rm obs}$ covariance matrix of experimental and theoretical
uncertainties, and $N_{\rm obs}$ is the number of observables.
We decompose the
theoretical predictions in the SMEFT into standard-model (SM)
contributions and beyond-the-standard-model (BSM) contributions as
\begin{equation}
O^{\rm SMEFT}=O^{\rm SM}+O^{\rm BSM},
\end{equation}
and rewrite $\chi^2$ as
\begin{equation}
\chi^2 = (O^{\rm diff}-O^{\rm BSM})^T (\sigma^2)^{-1}
(O^{\rm diff}-O^{\rm BSM}),
\label{eq-chisq-2}
\end{equation}
where 
\begin{equation}
O^{\rm diff}=O^{\rm exp}-O^{\rm SM}.
\end{equation}

Now we wish to put $\chi^2$ in Eq.~(\ref{eq-chisq-2}) into the 
linear-least-squares form. First, since the covariance matrix is 
symmetric, we can diagonalize it:
\begin{equation}
U^{-1}_{\rm cov} \sigma^2 U_{\rm cov} = \hat \sigma^2.
\end{equation}
We note that the diagonal matrix $\hat \sigma^2$ and the unitary matrix 
$U_{\rm exp}$ can be found conveniently 
from the SVD decomposition of $\sigma^2$, although other diagonalization 
methods could also be used. Then, we can write $\chi^2$ as
\begin{equation}
\chi^2 = (O^{\rm diff}-O^{\rm BSM})^T U_{\rm cov}(\hat\sigma^2)^{-1}
U_{\rm cov}^{-1}(O^{\rm diff}-O^{\rm BSM}).
\label{eq-chisq-3}
\end{equation}

Since the diagonal matrix $\hat \sigma^2$ is positive definite, the quantity 
$(\hat \sigma^2)^{-\frac{1}{2}}$ is well defined. Therefore, we can 
normalize the observables in the new basis to unit error by writing
\begin{subequations}
\begin{eqnarray}
\hat{O}^{\rm diff} &=& (\hat \sigma^2)^{-\frac{1}{2}} U_{\rm cov}^{-1} 
O^{\rm diff}, 
\\
\hat{O}^{\rm BSM}  &=& (\hat \sigma^2)^{-\frac{1}{2}} U_{\rm cov}^{-1}
  O^{\rm BSM}.
\end{eqnarray}
\end{subequations}
Now $\chi^2$ has the form
\begin{equation}
\label{eq:chisq_diagonal}%
\chi^2 = (\hat{O}^{\rm BSM} - \hat{O}^{\rm diff})^2. 
\end{equation}
Since $O^{\rm BSM}$ is linear in the SMEFT coefficients, we can write it
in the form 
\begin{equation}
O^{\rm BSM}= HC, 
\end{equation}
where $C$ is an $N_{\rm coeff}$-dimensional column vector of SMEFT
coefficients, $H$ is an $N_{\rm obs} \times N_{\rm
coeff}$ matrix, and $N_{\rm coeff}$ is the number of SMEFT coefficients. 
It follows that 
\begin{equation}
\hat{O}^{\rm BSM}=M C,
\end{equation}
where 
\begin{equation}
M=(\hat{\sigma}^2)^{-1}U_{\rm cov}^{-1} H.
\end{equation}
Hence, in order to constrain the SMEFT coefficients, we
minimize
\begin{equation}
\chi^2 = (MC - \hat{O}^{\rm diff})^2, 
\end{equation}
which is a linear-least-squares problem. As was described in
Sec.~\ref{sec:SVD}, the solution of this minimization problem can be
obtained from the SVD decomposition $M= UWV^\dag$:
\begin{equation}
\bar{C} = V W^{-1} U^\dag \hat{O}^{\rm diff}. 
\label{eq:min-solution}
\end{equation}
We emphasize that the expression in Eq.~(\ref{eq:min-solution}) provides
a convenient way to obtain the values of the coefficients that are well
determined, even in the presence of flat directions in the coefficient
space. If there are flat directions, then the formal solution of the linear
least squares problem\footnote{See, for example, Eq.~(4.3) of
Ref.~\cite{Aoude:2020dwv}.}
\begin{equation}
\label{eq:formal-solution}
\bar{C}=\left[H^T(\sigma^2)^{-1}H \right]^{-1}
H^T (\sigma^2)^{-1} O_{\rm diff}
\end{equation}
is ill defined because the ordinary inverse of 
the SMEFT covariance matrix
\begin{equation}
\sigma_{\rm SMEFT}^2=H^T(\sigma^2)^{-1}H
\label{eq:covariance-matrix}
\end{equation}
does not exist.
However, in Eq.~(\ref{eq:min-solution}), $W^{-1}$ is the Moore-Penrose
pseudoinverse of $W$ and is well defined, even in the presence of flat
directions. Owing to the properties of the pseusdoinverse, the
expression in Eq.~(\ref{eq:min-solution}) sets the values of the
undetermined coefficients to zero. 

As we will demonstrate, the best-fit values of the individual SMEFT
coefficients and their uncertainties do not accurately characterize the
constraints on the SMEFT coefficients. The principal components in the
coefficient space and their uncertainties parametrize the best fit in a
much more meaningful form.  As we have mentioned, each element of
$P^C=V^\dagger C$ gives one of the principal components of the
SMEFT coefficients. The best-fit values of the principal
components are given by the elements of $\bar{P}^C$ in
Eq.~(\ref{eq:best-fit-values}), and the one-standard-deviation
uncertainty on each principal component is given by the
corresponding element of $\Delta \bar{P}^C$ in
Eq.~(\ref{eq:uncertainty-formula}).

In terms of the SVD quantities, the covariance matrix of the SMEFT
coefficients is obtained by using $V$ to rotate $W$ back to the original
basis of SMEFT coefficients:
\begin{equation}
\sigma^2_{\rm SMEFT}=VW^{-1}(W^{-1})^\dagger V^\dagger.
\label{eq:smeft-covariance}
\end{equation}
As is standard, the covariance matrix for the situation in which one has
marginalized over some of the coefficients is obtained by striking from
the full covariance matrix the rows and columns that correspond to the
marginalized coefficients \cite{schon-lindsten}. Although the covariance
matrix contains the same information as the uncertainties in the
principal components, we will see that the presentation of uncertainties
in principal-component form leads to a clearer picture of the
constraints on the SMEFT coefficients.

The Fisher information matrix is given by the inverse of the
covariance matrix in Eq.~(\ref{eq:covariance-matrix}):
\begin{equation}
{\cal I}=\sigma_{\rm SMEFT}^2= H^T (\sigma^2)^{-1} H,
\end{equation}
from which it follows, using Eq.~(\ref{eq:smeft-covariance}), that 
\begin{equation}
{\cal I}=VW^\dagger W V^\dagger.
\end{equation}
In Refs.~\cite{Berthier:2015gja,
Berthier:2016tkq,Brivio:2017bnu,Brehmer:2017lrt,Brehmer:2019xox,Aoude:2020dwv},
the principal components of the SMEFT coefficients ($P^C$) are obtained by
diagonalizing the Fisher matrix. We note that the Fisher information is
the absolute square of the matrix $M=UWV^\dagger$ that is
analyzed in the SVD approach. Indeed, one method for implementing SVD
involves finding the eigenvectors and eigenvalues of $W$ by
diagonalizing the absolute square of $M$.  However,
it is well known that this method can be imprecise. Other, more
precise, methods for carrying out SVD have been devised~\cite{kalman-1996}. 
The reason for the
possible imprecision in diagonalizing the Fisher matrix is that
the absolute square of $M$ has a condition
number that is equal to the square of the condition number of
$M$. If the condition number is very large,
as is the case when there are nearly flat directions such
that $M$ has eigenvalues that differ
greatly in size, then there could, in principle be practical
advantages in using SVD, rather than diagonalization of the Fisher
matrix, to obtain the principal components.
      
A conceptual advantage of the SVD method over the Fisher-matrix method
is that it relates the principal components of the SMEFT
coefficients ($P^C$) to the principal components of the observables
($P^{O}$). Because $W$ is diagonal, the relationship is
one-to-one: 
\begin{equation}
P^C_i=W_{ii} P^{O}_i.
\label{eq:PC-relation}
\end{equation}
The relation in Eq.~(\ref{eq:PC-relation}) shows which linear
combination of observables affects a given
SMEFT-coefficient principal component. This insight could allow one to
identify new measurements that could further constrain a
poorly constrained SMEFT-coefficient principal component. Conversely,
the relation in Eq.~(\ref{eq:PC-relation}) would allow one to identify a
linear combination of observables that would be particularly sensitive
to the existence of physics beyond the standard model, as parametrized by a
SMEFT-coefficient principal component.

\section{Application to top-quark decay \label{sec:top-decay}}

In this section, we illustrate the PCA/SVD method for fitting the SMEFT
coefficients by applying it to the case of top-quark decay to a $b$
quark and a $W$ boson.

\subsection{SMEFT operators}

We work in the Warsaw-basis \cite{Buchmuller:1985jz} of SMEFT operators, 
and our notation is similar to that of Ref.~\cite{Dedes:2017zog}.
Following Ref.~\cite{Boughezal:2019xpp}, we fit the coefficients 
$C_{tW}$, $C_{bW}$, $C_{\phi tb}$, $C_{tg}$, and $C_{bg}$, which
correspond to the operators
\begin{eqnarray}
{\cal O}_{tW}&=&{\cal O}_{\substack{uW\\33}},\nonumber\\
{\cal O}_{bW}&=&{\cal O}_{\substack{dW\\33}},\nonumber\\
{\cal O}_{\phi tb}&=&{\cal O}_{\substack{\phi ud\\33}},\nonumber\\
{\cal O}_{tg}&=&{\cal O}_{\substack{ug\\33}},\nonumber\\
{\cal O}_{bg}&=&{\cal O}_{\substack{dg\\33}},
\end{eqnarray}
with
\begin{eqnarray}
{\cal O}_{\substack{uW\\pr}}&=&\bar q_p\sigma^{\mu\nu}u_r \tau^I \tilde\phi 
     W_{\mu\nu}^I,\nonumber\\
{\cal O}_{\substack{dW\\pr}}&=&\bar q_p\sigma^{\mu\nu}d_r  \tau^I \phi
     W_{\mu\nu}^I,\nonumber\\
{\cal O}_{\substack{\phi ud\\pr}}&=&i(\tilde\phi^\dagger D_\mu \phi)
     (\bar u_p \gamma^\mu d_r),\nonumber\\
{\cal O}_{\substack{uq\\pr}}&=&\bar q_p\sigma^{\mu\nu}T^A u_r \tilde\phi  
      G_{\mu\nu}^A, \nonumber\\
{\cal O}_{\substack{dg\\pr}}&=&\bar q_p\sigma^{\mu\nu}T^A d_r \tilde\phi  
      G_{\mu\nu}^A.
\end{eqnarray}
We also consider the coefficients $C_{qq}^{(1)}$, $C_{qq}^{(3)}$,
$C_{qu}^{(1)}$, $C_{qu}^{(8)}$ and $C_{lq}^{(3)}$, which correspond to
the four-fermion operators
\begin{eqnarray}
{\cal O}_{qq}^{(1)}&=&(\bar q_p\gamma^\mu q_r)(\bar q_s\gamma_\mu q_t),
\nonumber\\
{\cal O}_{qq}^{(3)}&=&(\bar q_p\gamma^\mu \tau^a q_r)
(\bar q_s\gamma_\mu \tau^a q_t),
\nonumber\\
{\cal O}_{qu}^{(1)}&=&(\bar q_p\gamma^\mu q_r)
(\bar u_s\gamma_\mu  u_t),
\nonumber\\
{\cal O}_{qu}^{(8)}&=&(\bar q_p\gamma^\mu T^Aq_r)
(\bar u_s\gamma_\mu  T^A u_t),
\nonumber\\
{\cal O}_{lq}^{(3)}&=&(\bar q_p\gamma^\mu \tau^a q_r)
(\bar q_s\gamma_\mu \tau^a q_t).
\end{eqnarray}
Here, $q_p$ ($l_p$) is a left-handed quark (lepton) isospin doublet with
generation index $p$, $u_r$ and $d_r$  are the up and down right-handed
isospin singlets with generation index r, $l_r$ is the lepton
right-handed isospin singlet with generation index r, $\phi$ is the Higgs
isospin doublet, $\tilde\phi=i\tau_2 \phi^*$ is the hypercharge-conjugate
Higgs doublet, $\tau$ is a Pauli matrix, $W_{\mu\nu}^I$ is the
field-strength tensor for the $SU(2)_I$ gauge bosons with isospin index
$I$, the $\gamma$'s are Dirac matrices,
$\sigma_{\mu\nu}=i[\gamma_\mu,\gamma_\nu]$, $G_{\mu\nu}^A$ is the gluon
field-strength tensor with color index $A$, and $T^A$ is a  color matrix
in the fundamental representation with color index $A$.

\subsection{Experimental inputs}

We take experimental values of the total top-quark decay rate and the 
helicity fractions from the 
Particle Data Group compilation~\cite{Tanabashi:2018oca}:
\begin{eqnarray}
\Gamma_{\rm tot}^{\rm exp}&=&1.41_{-0.15}^{+0.19}~{\rm GeV},\nonumber\\
F_L^{\rm exp}&=&0.687\pm 0.018,\nonumber\\
F_-^{\rm exp}&=&0.320\pm 0.013.
\end{eqnarray}
In our analysis, we symmetrize the uncertainties in $\Gamma_{\rm 
tot}^{\rm exp}$ by shifting the central value. That is, we take 
$\Gamma_{\rm tot}^{\rm exp}= 1.43\pm 0.17~{\rm GeV}$.
The correlation matrix of the experimental uncertainties is  
\cite{Khachatryan:2016fky} 
\begin{equation}
\rho = \begin{pmatrix}
1.0 & 0 & 0 \\
0 & 1.0 & -0.87 \\
0 & -0.89 & 1.0
\end{pmatrix}.
\end{equation}
Then, the experimental inputs for our SVD analysis are
$O^{\rm exp} = (\Gamma_{\rm tot}^{\rm exp}, F_L^{\rm exp}, F_-^{\rm exp})^T$ 
and $\sigma^2_{ij} = \sigma_i \rho_{ij} \sigma_j$. Here $i,j = 1,2$, and 
$3$ correspond to $\Gamma_{\rm tot}^{\rm exp}$,
$F_L^{\rm exp}$, and $F_-^{\rm exp}$, respectively, and
the $\sigma_i$ are the experimental uncertainties.

\subsection{Theoretical inputs}

We make use of the expressions in Ref.~\cite{Boughezal:2019xpp} for
SMEFT contributions to $\Gamma_{\rm tot}$, the total decay width to $b
W$, $F_L$, the fractional decay rate for a longitudinally polarized $W$
boson, and $F_-$, the fractional decay rate for a $W$ boson with
negative helicity.\footnote{The expressions for these quantities in
the published version of Ref.~\cite{Boughezal:2019xpp} 
have been corrected in the second
arXiv version of that paper~\cite{wiegand}. } We include
the QCD corrections that are given in Ref.~\cite{Boughezal:2019xpp}. We
note that the standard-model QCD corrections are also given in
Ref.~\cite{Fischer:2000kx} and that an analysis of the SMEFT
contributions to $t$ quark decay has also been given in
Ref.~\cite{Zhang:2014rja}.

For purposes of this demonstration, we do not include uncertainties in
the theoretical predictions. They could be incorporated into the
analysis by adding the theoretical covariance matrix to the experimental
covariance matrix.\footnote{If the theoretical and experimental
uncertainties are correlated, say, through the use of a common parameter
in the theoretical and experimental analyses, then one would need to
construct a complete covariance matrix of theoretical and experimental
uncertainties, including entries for that parameter, and then
marginalize over that parameter.}

The input parameters for the theoretical calculation are given in 
Table~\ref{tab:parameters} and are identical to those in 
Ref.~\cite{Boughezal:2019xpp}, except that we evaluate $\alpha_s$ at the 
scale $m_t$, rather than the scale $M_Z$.
\begin{table}[h!]
\centering                                                          
\begin{tabular}{cccc}
\hline\hline                       
 $M_Z$ & $91.1876\,\textrm{GeV}$ & $M_W$ & $80.379\,\textrm{GeV}$  \\
 $v$& $246 \,\textrm{GeV}$ & $m_t$ & $173.0 \,\textrm{GeV} $ \\
 $m_b$& $4.78  \,\textrm{GeV}$ & $G_F$ & $1.1664 \times 10^{-5} \,       
  \textrm{GeV}^{-2}$  \\
 $\alpha^{-1}_\textrm{em}$ & $137.036$ & $\alpha_s(m_t)$ & $0.1081$ \\
\hline\hline
\end{tabular}
\caption{Input parameters for the theoretical 
calculation.\label{tab:parameters}}
\end{table}
We compute the electroweak coupling $\bar g$ from \cite{Dawson:2019clf}
\begin{equation}
\bar g^2=2\sqrt{2} G_F M_Z^2\left(1+\sqrt{1-\frac{4\pi \alpha_{\rm em}}
{\sqrt{2} G_F M_Z^2}}\right).
\end{equation}
We set the SMEFT cutoff to be $\Lambda=500$~GeV.

\subsection{Fit with one SMEFT coefficient}

In Table~\ref{tab:single-coefficient}, we show the best-fit values of
the SMEFT coefficients and their two-standard-deviation uncertainties
that are obtained by setting all of the coefficients to zero, except
for one. This is a widely used approach for constraining the SMEFT
coefficients. However, as we will see, it can be quite misleading.
\begin{table}[h!]
\centering
\begin{tabular}{c|c|c|c}
\hline\hline
$C_{tW}$ &       $0.0644 \pm 0.100$ &  $C_{qq}^{(1)}$ & $-4.81  \pm 37.1$ \\
$C_{tg}$ &       $-3.65   \pm 8.22$ &   $C_{qq}^{(3)}$ & $0.656  \pm 5.07$ \\
$C_{bW}$ &       $0.679  \pm 1.03$ &   $C_{qu}^{(1)}$ & $14.1   \pm 56.4$ \\
$C_{bg}$ &       $-13.5  \pm 17.9$ &   $C_{qu}^{(8)}$ & $10.6   \pm 42.3$ \\
$C_{\phi tb}$ &  $2.94   \pm 5.56$ &   $C_{lq}^{(3)}$ & $-4.35  \pm 33.6$ \\
\hline \hline
\end{tabular}
\caption{Best-fit values of the SMEFT coefficients and their
two-standard-deviation uncertainties computed by setting all of the
coefficients to zero, except for one. \label{tab:single-coefficient}}
\end{table}

\subsection{Fit with three SMEFT coefficients}

Now let us consider
the case in which only the first three coefficients 
in Table~\ref{tab:single-coefficient} are nonzero. Then, we can compute 
the best-fit values of those coefficients and their uncertainties, marginalized 
over the other two coefficients. As we explained earlier, the latter can 
be obtained from the diagonal values of the covariance matrix. The 
results of this computation are shown in 
Table~\ref{tab:3-coeffs-marginalized}. 
\begin{table}
\begin{tabular}{c|c}
\hline\hline
$C_{tW}$ & $0.0209 \pm 0.333$
\\
$C_{tg}$ & $4.73 \pm 62.8$
\\
$C_{bW}$ & $1.13 \pm 8.72$
\\
\hline \hline
\end{tabular}
\caption{Best-fit        
values of the SMEFT coefficients obtained by keeping only three 
coefficients nonzero. The uncertainties are the two-standard-deviation  
uncertainties that are obtained by marginalizing over two of the coefficients.
\label{tab:3-coeffs-marginalized}}
\end{table}
As can be seen, the central values have shifted substantially relative
to those in Table~\ref{tab:single-coefficient}, and the uncertainties
have increased, in some cases by almost an order of magnitude. Clearly,
the single-coefficient values and uncertainties in
Table~\ref{tab:single-coefficient} are not indicative of the true
constraints on the SMEFT coefficients in the presence of three nonzero
coefficients. However, the large uncertainties in
Table~\ref{tab:3-coeffs-marginalized} paint an unduly pessimistic
picture of the constraints that can be achieved.

In order to see this, let us 
carry out a PCA with three nonzero coefficients. 
Making use of the SVD method, we obtain 
\begin{subequations}%
\begin{equation}
U^\dagger=\left(\begin{array}{ccc}
-0.273& -0.770& -0.576\\
-0.276& -0.511& 0.814\\
0.922& -0.381& 0.0732
\end{array}
\right),
\end{equation}
\begin{equation}
V^\dag=\left(\begin{array}{ccc}
-0.999& 0.000697 & -0.0414\\
0.0411& 0.136 & -0.990 \\
-0.00496& 0.991& 0.136
\end{array}
\right),
\end{equation}
and
\begin{equation}
W=\left(\begin{array}{ccc}
19.9& 0 & 0\\
0& 1.77& 0\\
0& 0& 0.0315
\end{array}
\right).
\end{equation}
\end{subequations}
Here, and throughout the remainder of this paper, when we present the
array $V^\dagger$, the columns correspond to the following order of the
SMEFT coefficients: $C_{tW}$, $C_{tg}$, $C_{bW}$, $C_{bg}$, $C_{\phi
tb}$, $C_{qq}^{(1)}$, $C_{qq}^{(3)}$, $C_{qu}^{(1)}$, $C_{qu}^{(8)}$,
$C_{lq}^{(3)}$. When we present the array $U^\dagger$, the columns
correspond to the following order of the experimental observables:
$\Gamma_{\rm tot}^{\rm exp}$, $F_L^{\rm exp}$, $F_-^{\rm exp}$.

From the expression for $V^\dagger$, we see that the principal
components of the SMEFT coefficients are 
\begin{eqnarray}
P^C_1&=&0.999 C_{tW}- 0.000697 C_{tg} +0.0414 C_{bW},\nonumber \\
P^C_2&=& -0.0411 C_{tW} -0.136 C_{tg} +0.990 C_{bW},\nonumber\\
P^C_3&=& -0.00496 C_{tW} +0.991 C_{tg} +0.136 C_{bW}.
\end{eqnarray}

The best-fit values of the principal components are given by the
elements of $\bar{P}^C$ in Eq.~(\ref{eq:best-fit-values}), and
their two-standard-deviation uncertainties are given by twice the
inverse of the corresponding diagonal element of $W$. The results are
\begin{eqnarray}
\label{eq:PA-errors-3}%
\bar{P}^C_1&=& -0.0645 \pm 0.100,\nonumber\\
\bar{P}^C_2&=& -0.477 \pm 1.13,\nonumber\\
\bar{P}^C_3&=& 4.84 \pm 63.4.
\end{eqnarray}
We see that $P^C_1$ and $P^C_2$ are much better constrained 
than any of the individual coefficients and that only $P^C_3$ is poorly 
constrained. The PCA analysis clearly allows one to 
access a much more powerful set of constraints than do the analyses of 
individual SMEFT coefficients. 

The principal components of the observables are given by the rows of
$U^\dagger$:
\begin{eqnarray}
P^{O}_1&=& -0.273 {O}_1- 0.770 {O}_2
-0.576 {O}_3, \nonumber\\
P^{O}_2&=& -0.276 {O}_1- 0.511 {O}_2 + 0.814
{O}_3, \nonumber\\
P^{O}_3&=&+0.992 {O}_1-0.381 {O}_2+0.0732
{O}_3 .
\end{eqnarray}
As we have mentioned, each observable principal component $P_i^O$ is
coupled through $W$ to a single coefficient principal component
$P_i^C$, and {\it vice versa}.
From the elements of $W$, we see that $P^C_1$ is highly constrained by
$P^{O}_1$ and, conversely, $P^{O}_1$ would be very sensitive
to new physics, as parametrized by $P^C_1$. This suggests that one might
look for new physics be tightening the experimental bound on $P^{O}_1$.  
On the other hand, $P^{O}_2$ is less strongly coupled to
$P^C_2$, and $P^{O}_3$ is even less strongly coupled to
$P^C_3$. This suggests that it might be useful to search for additional
SMEFT operators that affect the linear combination of obervables in
$P^{O}_2$ and $P^{O}_3$.

\subsection{Fit with five SMEFT coefficients}

Now suppose that we keep only the first five SMEFT coefficients in 
Table~\ref{tab:single-coefficient} nonzero. In this case, we have more 
SMEFT coefficients than observables, and so the individual coefficients 
cannot be fit unambiguously. Furthermore, because there are necessarily 
flat directions, the marginalization over some sets of SMEFT coefficients is 
ill-defined. Nevertheless, the PCA/SVD approach allows us to find meaningful 
constraints. We obtain
\begin{equation}
U^\dagger=\left(\begin{array}{ccc}
-0.273& -0.770& -0.576\\
0.275& 0.512& -0.814\\
-0.922& 0.381& -0.0720
\end{array}
\right),
\end{equation}
\begin{equation}
V^\dag =\left(\begin{array}{ccccc}
-0.999& 0.000698& -0.0414& 0.00390& -0.00316 \\
0.0411& 0.134& -0.969& 0.0443& -0.196 \\
0.00282& -0.931& -0.0541& 0.0222& -0.361 \\
0.00199& 0.0163& 0.0453& 0.999& 0.0128 \\
0.00647& -0.340& -0.231& 0.00431& 0.912 \\
\end{array}
\right),
\end{equation}
and 
\begin{equation}                                                      
W=\left(\begin{array}{ccccc}  
19.9& 0& 0& 0& 0\\
0& 1.80& 0& 0& 0\\
0& 0& 0.0339& 0& 0
\end{array}
\right).
\end{equation}
The coefficients of the SMEFT coefficients of a given principal
component are given by the entries in the corresponding row in
$V^\dagger$. The best-fit values for the first three 
principal components and their two-standard-deviation uncertainties are 
\begin{eqnarray}
\label{eq:PA-errors-5}%
\bar{P}_1^C&=& -0.0645 \pm 0.100,\nonumber \\
\bar{P}_2^C&=& -0.467 \pm 1.11,\nonumber \\
\bar{P}_3^C&=& -4.46 \pm 59.0. 
\end{eqnarray}
The linear combinations of SMEFT coefficients that are unconstrained
have been sequestered by the SVD procedure. Comparing
Eq.~(\ref{eq:PA-errors-5}) with Eq.~(\ref{eq:PA-errors-3}), we see that
the best-fit values and uncertainties in the first two principal
components are remarkably stable as new SMEFT contributions are
introduced. This reflects the fact that the observables are relatively
insensitive to the SMEFT contributions that are proportional to $C_{bg}$
and $C_{\phi tb}$, as can be seen from the small coefficients of
$C_{bg}$ and $C_{\phi tb}$ in the first two principal components.  We see
that the principal components of the observables, which are given by the
rows of $U^\dag$, are almost unchanged in comparison with those
from the three-coefficient fit.

\subsection{Fit with ten SMEFT coefficients \label{sec:10-coeff-fit}}

Next we apply the SVD method to the complete set of ten SMEFT coefficients in 
Table~\ref{tab:single-coefficient}. 
We list only the three principal 
components that are constrained. They are
\begin{eqnarray}
P_1^C&=&-0.999 C_{tW}+ 0.000698 C_{tg} -0.0414 C_{bW}+ 0.00390 
C_{bg} -0.00316 C_{\phi tb}\nonumber\\ 
&&\qquad +0.000737 C_{qq}^{(1)} -0.00540 
C_{qq}^{(3)} -0.000570 C_{qu}^{(1)} -0.000760 C_{qu}^{(8)} + 
0.000815 C_{lq}^{(3)},\nonumber\\
P_2^C&=&0.0406 C_{tW}+ 0.134 C_{tg} -0.967 C_{bW}+ 0.0442
C_{bg} -0.196 C_{\phi tb} \nonumber\\
&&\qquad  -0.00856 C_{qq}^{(1)}
+0.0628 C_{qq}^{(3)} -0.00550 C_{qu}^{(1)} -0.00733 C_{qu}^{(8)}  
-0.00947 C_{lq}^{(3)},\nonumber\\
P_3^C&=&0.00819 C_{tW} -0.0753 C_{tg} -0.0672 C_{bW}+ 0.00484
C_{bg} -0.0452 C_{\phi tb} \nonumber\\
&&\qquad + 0.131 C_{qq}^{(1)} -0.964
C_{qq}^{(3)} +0.0854 C_{qu}^{(1)} +0.114  C_{qu}^{(8)}  
+ 0.145 C_{lq}^{(3)}.
\end{eqnarray}
The best-fit values and two-standard-deviation uncertainties for these 
principal components are 
\begin{eqnarray}
\label{eq:10-coeff-results}
\bar{P}_1^C &=&-0.0645\pm 0.100,\nonumber\\
\bar{P}_2^C &=&-0.465\pm 1.11,\nonumber\\
\bar{P}_3^C &=&-0.432\pm 5.32.
\end{eqnarray}
As can be seen, the first two principal components remain quite stable 
in best-fit value and uncertainty as additional SMEFT coefficients are 
introduced, reflecting the relative insensitivity of the observables to 
the additional SMEFT coefficients.

The principal components of the observables can be obtained from $U^\dag$,
which is given by 
\begin{equation}
U^\dagger=\left(\begin{array}{ccc}
-0.273& -0.770& -0.576\\
0.287& 0.507& -0.813\\
-0.918& 0.387& -0.0829
\end{array}
\right).
\end{equation}
This result for $U^\dagger$ is not very different from the results
from the three- or five-coefficient fits.

The results in Eq.~(\ref{eq:10-coeff-results}) express precisely the
constraints on the 10-coefficient fit that follow from the input
top-quark decay rates. As such, they should be considered to
supersede the results from the one- and two-coefficient fits that
were given in Ref.~\cite{Boughezal:2019xpp}, since those fits fail
to account for the high degree of correlation between uncertainties
in the SMEFT coefficients, and, consequently, are quite
misleading.

\subsection{Fit with a flat direction in coefficient space}

Next, we examine the case in which the number of SMEFT coefficients
and the number of observables are equal, but there is a hidden flat
direction. In order to construct an example of this situation, we keep
three SMEFT coefficients, $C_{tW}$, $C_{tg}$, and $C_{bW}$, nonzero and
set the remaining SMEFT coefficients to zero. Let $a_{tW}$, $a_{tg}$,
and $a_{bW}$ be the coefficients of $C_{tW}$, $C_{tg}$, and $C_{bW}$ in
$O^{\rm SMEFT}$. Then, the following replacement creates an approximate
artificial flat direction in the space of $C_{tg}$ and $C_{bW}$:
\begin{equation}
a_{tg}\to \epsilon a_{tg}+(1-\epsilon)r a_{bW}.
\label{eq:flat-sub}
\end{equation}
In the limit $\epsilon\to 0$, there is an exact flat direction in the
space of SMEFT coefficients. 

As a numerical example, we take scale factor $r$ in
Eq.~(\ref{eq:flat-sub}) to be $-3.2$. As $\epsilon$ approaches zero,
conventional fitting procedures that use gradients of $\chi^2$ to find a
minimum in $\chi^2$ have numerical difficulties. Let us consider, for
example, the situation for $\epsilon=10^{-6}$. The {\it Mathematica} routine
FindMinimum can be used to minimize the $\chi^2$. This routine comes
with a number of options for the method to be used in finding the
minimum. Using {\it Mathematica} version 11.3~\cite{mathematica-11}, we find 
that the conjugate-gradient method algorithm yields
\begin{eqnarray}
C_{tW}&=&0.0445,\nonumber\\
C_{tg}&=&0.236,\nonumber\\
C_{bW}&=&1.24,
\end{eqnarray}
Newton's method yields 
\begin{eqnarray}
C_{tW}&=&0.0410,\nonumber\\
C_{tg}&=&7.00\times 10^5,\nonumber\\
C_{bW}&=&2.24\times 10^6, 
\end{eqnarray}
and the principal-axis method yields 
\begin{eqnarray}
C_{tW}&=&0.0445,\nonumber\\
C_{tg}&=&0.161,\nonumber\\
C_{bW}&=&1.00. 
\end{eqnarray}
Owing to the existence of a nearly flat direction, the results vary
wildly, depending on the algorithm that is used in minimizing $\chi^2$.
While this behavior is to be expected in under-determined problems, it is
difficult to draw any conclusion from such results. In contrast,
meaningful constraints can be found by using PCA. From SVD, we obtain
\begin{eqnarray}
\label{eq:SVD_nearflat_V}%
V^\dag = 
\left(
\begin{array}{ccc}
 -0.988845 & 0.142167 & -0.0444274 \\
 -0.148947 & -0.943833 & 0.294948 \\
 -1.49286\times 10^{-9} & 0.298275 & 0.95448 \\
\end{array}
\right)  
\end{eqnarray}
and 
\begin{eqnarray}
\label{eq:SVD_nearflat_W}%
W = 
\left(
\begin{array}{ccc}
 20.1307 & 0 & 0 \\
 0 & 5.80765 & 0 \\
 0 & 0 & 9.49318\times 10^{-9} \\
\end{array}
\right).
\end{eqnarray}
The coefficient principal components are 
\begin{eqnarray}
\label{eq:princ-comps-flat}%
P_1^C &=& -0.989 C_{tW} + 0.142 C_{tg} -0.0444 C_{bW},\nonumber\\
P_2^C &=& - 0.149 C_{tW} - 0.944 C_{tg} + 0.295 C_{bW},\nonumber\\ 
P_3^C &=& -1.49\times 10^{-9} C_{tW} + 0.298 C_{tg} + 0.954 C_{bW}.
\end{eqnarray}
From the result for $W$, it follows that 
$P_1$ and $P_2$ are well constrained, while $P_3$ is not, as is
evident from the near-vanishing of the corresponding diagonal value of
$W$.\footnote{We keep more significant digits than usual in  
the results for $V^\dag$ and $W$ in
Eqs.~(\ref{eq:SVD_nearflat_V}) and~(\ref{eq:SVD_nearflat_W}) for 
purposes of later comparison with the results for $V^\dag$ and $W$ when 
$\epsilon=0$.}
The observable principal component $P_3^O$ has essentially no
sensitivity to the physics beyond the standard model that is embodied in
$P_3^C$.
We see that the SVD/PCA method has constrained
the principal components that contain a contribution that is
proportional to $-3.2 C_{tg}+C_{bW}$ and has identified as unconstrained
the principal component that contains a contribution that is
proportional to $C_{tg}+3.2C_{bW}$, which corresponds to the flat
direction. The best-fit values of the principal components and their
two-standard deviation uncertainties are
\begin{eqnarray}
\label{eq:princ-comps-flat-bestfit}%
\bar{P}_1^C &=& -0.0655 \pm 0.0994,\nonumber\\
\bar{P}_2^C &=& 0.136 \pm 0.344,\nonumber\\
\bar{P}_3^C &=& 1.58\times 10^7 \pm 2.11\times10^8.
\end{eqnarray}
We can invert the relations in Eq.~(\ref{eq:princ-comps-flat}), 
using the rows of $V$ to obtain the coefficients. The result is
\begin{eqnarray}
\label{eq:inverse-princ-comps-flat}%
C_{tW} &=& -0.989 P_1^C - 0.149 P_2^C -1.49\times 10^{-9} P_3^C,\nonumber\\
C_{tg} &=&  0.142 P_1^C -0.944 P_2^C +0.298 P_3^C,\nonumber\\ 
C_{bW} &=& -0.0444 P_1^C +0.295 P_2^C +0.954 P_3^C,
\end{eqnarray}
From Eq.~(\ref{eq:inverse-princ-comps-flat}), it is easily seen that, 
to good approximation, the differences between the three results from 
FindMinimum correspond to differences in the value of $P_3^C$.

When $\epsilon=0$ and there is an exact flat direction, numerical
minimization of $\chi^2$ with respect to the SMEFT coefficients
would fail to converge to a result. However, the SVD method still yields
meaningful constraints. Specifically, we have
\begin{equation}
V^\dagger=\left(\begin{array}{ccc}
-0.988845& 0.142168& -0.0444274\\
-0.148948& -0.943833& 0.294948\\
1.29707\times 10^{-17}   & 0.298275& 0.95448
\end{array}\right)
\end{equation}
and
\begin{equation}                                       
W=\left(\begin{array}{ccc}                     
20.1307& 0&0\\                       
0& 5.80765& 0\\                       
0& 0& 0                                   
\end{array}\right),
 \end{equation}
which are very close to the results for $\epsilon=10^{-6}$.

\subsection{Comparison of SVD with diagonalization of the Fisher
  matrix}

Finally, let us compare the speed and accuracy of the computation of the
eigenvalues and eigenvectors of the Fisher matrix through direct
calculation and through the use of SVD. We consider the case of ten
SMEFT coefficients, which is the most challenging computationally. In
that case, when we evaluate expressions in
{\it Mathematica}~12~\cite{mathematica-12}, the direct-diagonalization and SVD
methods lead to eigenvalues of $W$ (square roots of the eigenvalues of
the Fisher matrix) whose relative differences are no more than $4\times
10^{-16}$ and eigenvectors whose nonzero components have relative
differences that are no more than $6.8\times 10^{-14}$. Clearly, these
are insignificant in comparison with other uncertainties in the fits,
although they might become more significant in fitting programs that
use single-precision arithmetic.  Differences in computation time
are also insignificant for matrices of this size, as computation times
for both methods are on the order of $10^{-5}$ seconds on a modern
laptop.

In the case of three nonzero coefficients with an artificial nearly flat
direction, the accuracy of the smallest eigenvalue suffers in the
case of the Fisher-matrix method, as it becomes negative ($-7\times
10^{-15}$ for the eigenvalue of the Fisher matrix versus $9\times
10^{-9}$ for the eigenvalue of $W$). However, this unphysical result
that arises from the Fisher-matrix method has no practical
consequences for a SMEFT analysis because large excursions of SMEFT
coefficients from zero can be bounded by appealing to power counting
arguments \cite{Berthier:2015gja}.

The issue of computation time might become more significant in the
analysis of larger matrices, such as those that would appear in a full
fit of the dimension-8 SMEFT coefficients. In particular, if the number
of SMEFT coefficients is much greater than the number of
observables, then the Fisher matrix is much larger than the matrix
that is analyzed in SVD. In this situation, the computation time for an
SVD analysis can be much less than for diagonalization of the Fisher
matrix.  For example, if one takes $M$ to be a random matrix $100\times
3000$ matrix, then the computation time to find eigenvalues and
eigenfunctions is about 1 second for SVD and 11 seconds for the
Fisher-matrix method in {\it Mathematica}~12 on a modern laptop .  These
computation-time differences would be magnified if many fits need be
performed, for example, in varying input parameters or in using the
iterative approach that is described in Sec.~\ref{sec:extension}. This
computational advantage of the SVD method over the Fisher-matrix
method arises only when the number of SMEFT coefficients is much
greater than the number of observables, and, in fact, the
Fisher-matrix method has the computational advantage when the
situation is reversed and the number of observables is much greater
than the number of SMEFT coefficients. That advantage would be
significant only when the number of observables is very large ($\agt
1000$).

\section{Extension to higher orders in the SMEFT 
expansion\label{sec:extension}}

The SVD method that we have presented is limited
to fitting problems in which the observables depend linearly on SMEFT
coefficients. This is the case for fits at the leading nontrivial order in
the effective-field-theory expansion, in which one considers only the
contributions of the interference of the dimension-6 SMEFT-operator
amplitudes with the dimension-4 standard-model-operator amplitudes. At
the next order in the SMEFT expansion, one would need to consider not
only the contribution from the square of the dimension-6 SMEFT-operator
amplitudes, but also the interference of the dimension-4
standard-model-operator amplitudes with the thousands of dimension-8
SMEFT-operator amplitudes---a task that is not likely to be undertaken
soon.

Nevertheless, beyond leading order, one might still apply the SVD method
by making use of an iterative procedure. One could first carry out a fit
that retains only the interference of the dimension-6 amplitude with the
dimension-4 amplitude. Then, one could compute the contribution of the
square of the dimension-6 amplitude, subtract it from the experimental
values of the observables, and carry out a new fit, including the
dimension-8 operators. This last step could be iterated to produce fits
of the desired accuracy. The iteration process should converge if the
effective-field-theory expansion is valid, that is, if the square of the
dimension-6 amplitude is less than the interference of the dimension-6
amplitude with the dimension-4 amplitude.

This method would yield best-fit values of the coefficients, but would
not give accurate results for the principal components. Instead, one
could compute the principal components as follows. First one could
obtain the Fisher matrix (inverse covariance matrix) by computing
analytically two derivatives of $\chi^2$ with respect to the SMEFT
coefficients and evaluating the result at the best-fit values of the
coefficients from the iterative procedure. The Fisher matrix could
be diagonalized by standard methods, and the principal components could
then be obtained from the elements of the unitary transformation that
effects the diagonalization. The uncertainties would be given by the
inverses of the square roots of the diagonal components of the Fisher
matrix. Although the Fisher-matrix method is used in the last
step, this step occurs only once in the procedure. The
``heavy-lifting,'' involving the repeated, iterative principal-component
analysis of large matrices, can be carried out by making use of SVD.

\section{Summary \label{sec:summary}}

In using experimental data to constrain the Wilson coefficients in
standard model effective field theory (SMEFT), a difficulty that often
arises is that observables may be insensitive to certain linear
combinations of SMEFT coefficients. That is, there may be ``flat
directions'' in the space of SMEFT coefficients. This difficulty can
arise because, in a partial analysis that is restricted to a particular
set of physical processes, the number of experimental observables may be
less than the number of SMEFT coefficients. In this case, it is clear
that some linear combinations of SMEFT coefficients would not be
constrained. However, it can happen that some linear combinations of
SMEFT coefficients are poorly constrained even when the number of
observables is equal to or greater than the number of SMEFT coefficients
to be fit.

A standard approach for dealing with this difficulty is to carry out a
principal-component analysis (PCA) of the SMEFT coefficients by
diagonalizing the Fisher information matrix
\cite{Berthier:2015gja,Berthier:2016tkq,Brivio:2017bnu,Brehmer:2017lrt,
  Brehmer:2019xox,Aoude:2020dwv}.  In this paper, we have presented an
alternative approach for carrying out the PCA that is based on
singular-value decomposition (SVD). As we have shown, the SVD method
provides information about the sensitivity of experimental
observables to SMEFT coefficients that is not accessible in the
Fisher-matrix method. That information could be used to
identify new measurements that could improve the constraint on a
poorly constrained SMEFT-coefficient principal component. It could
also be used to target particular linear combinations of observables
in searches for new physics.

In principle, SVD may offer superior precision in comparison with
diagonalization of the Fisher matrix because SVD involves the analysis
of a matrix that is better conditioned than the Fisher matrix. The
condition number of the Fisher matrix is the square of the condition
number of the matrix that is analyzed in the SVD method, and it
may become large if there are nearly flat directions in the space of
SMEFT coefficients. In practice, very small eigenvalues of the
Fisher that are associated with nearly flat directions are of little
physical consequence because the excursions of the corresponding
coefficient eigenvectors are limited by SMEFT power-counting
arguments~\cite{Berthier:2015gja}.

In the situation in which the number of SMEFT coefficients is very
large and the number of observables is much smaller, as might occur in
an analysis of dimension-8 SMEFT coefficients, the matrix that is
analyzed in the SVD method is much smaller than the Fisher matrix, and
the SVD method may accrue advantages in computational speed.

As we have pointed out, the SVD method also provides a convenient way to
compute the central values of the SMEFT coefficients for constrained
directions, even in the presence of flat directions.

We have demonstrated the application of the SVD method to the process
of top-quark decay to a $W$ boson and a $b$ quark. In this
demonstration, we give specific illustrations of the pitfalls in two
widely used fitting approaches, namely, (1) setting all of the
coefficients to zero except for one and (2) marginalizing over all
of the coefficients except for one. We show that approach (1) leads
to shifted central values of the coefficients and underestimated
uncertainties, while approach (2) leads to an overly pessimistic
assessment of uncertainties, which can be ameliorated through the use
of PCA. Our fit for the case of ten SMEFT coefficients should be
considered to supersede both the one- and two-coefficient fits in
Ref.~\cite{Boughezal:2019xpp}, which do not account for the
highly-correlated uncertainties in the SMEFT coefficients.

In the example of top-quark decay to a $W$ boson and a $b$ quark, the
matrices involved are of too small to reveal the putative
computational advantages of the SVD method. However, the advantages
might become significant in fits involving the many dimension-8 SMEFT
coefficients if multiple fits were needed, for example, in varying input
parameters or in using the iterative fitting procedure that we have
described.
 
Although the method that we have presented is limited to the case in
which the observables depend linearly on the SMEFT coefficients, we have
outlined in Sec.~\ref{sec:extension} an iterative extension of the
method that can be applied to the nonlinear situation, provided that
the contributions of the nonlinear terms to the observables are small
in comparison with the contributions of the linear terms. 
This is the case if the SMEFT expansion converges.

Finally, the method that we have presented relies on a Gaussian
probability analysis. While one might ultimately want to improve on
a Gaussian approach, it should certainly be adequate for the purpose of
carrying out exploratory studies in SMEFT.

\begin{acknowledgments}

We are grateful to Christopher Murphy for pointing out 
to us Ref.~\cite{Murphy:2017omb}, as well as several recent papers on 
global fitting of SMEFT coefficients.
We thank Radja Boughezal, Chien-Yi Chen, Frank Petriello, and Daniel
Wiegand for providing us with a version of Ref.~\cite{Boughezal:2019xpp}
that contains corrected expressions for the total width and helicity
fractions in top-quark decay to a $b$ quark and a $W$ boson. We also
thank Xiang-peng Wang for checking some of our {\it Mathematica} expressions.
The work of G.T.B.\ is supported by the U.S.\ Department of Energy,
Division of High Energy Physics, under Contract No.\ DE-AC02-06CH11357.
The work of H.S.C\ is funded by the Deutsche Forschungsgemeinschaft (DFG,
German Research Foundation) under Germany's Excellence Strategy -- EXC-2094 --
390783311. The submitted
manuscript has been created in part by UChicago Argonne, LLC, Operator
of Argonne National Laboratory. Argonne, a U.S.\ Department of Energy
Office of Science laboratory, is operated under Contract No.\
DE-AC02-06CH11357. The U.S.\ Government retains for itself, and others
acting on its behalf, a paid-up nonexclusive, irrevocable worldwide
license in said article to reproduce, prepare derivative works,
distribute copies to the public, and perform publicly and display
publicly, by or on behalf of the Government.

\end{acknowledgments}


\end{document}